\documentclass[lotsofwhite]{patmorin}
\usepackage{charter}
\usepackage{amsthm}
\usepackage{graphicx}

\newcommand{\comment}[1]{}

\newcommand{\seclabel}[1]{\label{sec:#1}}

\newcommand{\secref}[1]{\mbox{Section~\ref{sec:#1}}}

\newcommand{\tablabel}[1]{\label{tab:#1}}

\newcommand{\tabref}[1]{Table~\ref{tab:#1}}

\newcommand{\figlabel}[1]{\label{fig:#1}}

\newcommand{\figref}[1]{\mbox{Fig.~\ref{fig:#1}}}

\newcommand{\eqref}[1]{(\ref{eq:#1})}

\newtheorem{thm}{Theorem}{\bfseries}{\itshape}
\newcommand{\thmlabel}[1]{\label{thm:#1}}
\newcommand{\thmref}[1]{Theorem~\ref{thm:#1}}

\newtheorem{lem}{Lemma}{\bfseries}{\itshape}
\newcommand{\lemlabel}[1]{\label{lem:#1}}
\newcommand{\lemref}[1]{Lemma~\ref{lem:#1}}

{\bfseries}{\itshape}

{\bfseries}{\itshape}

\newtheorem{assumption}{Assumption}{\bfseries}{\rm}

\newcommand{\floor}[1]{\left\lfloor #1 \right\rfloor}

\newcommand{\email}[1]{\texttt{#1}}
\title{\MakeUppercase{Range Mode and Range Median Queries on Lists and Trees}%
	\thanks{This work was partly funded by the Natural Sciences
	and Engineering Research Council of Canada.}}
\author{Danny Krizanc\thanks{%
	Wesleyan University, 
	\email{dkrizanc@wesleyan.edu}} \and 
	Pat Morin\thanks{%
	Carleton University, \email{\{morin,michiel\}@cs.carleton.ca}} \and 
	Michiel Smid\footnotemark[2]}
\date{}

\begin{document}
\maketitle

\begin{abstract}
We consider algorithms for preprocessing labelled lists and trees so
that, for any two nodes $u$ and $v$ we can answer queries of the form:
What is the mode or median label in the sequence of labels on the path from
$u$ to $v$.
\end{abstract}

\section{Introduction}

Let $A=a_1,\ldots,a_n$ be a list of elements of some data type.  Many
researchers have considered the problem of preprocessing $A$ to answer
\emph{range queries}.  These queries take two indices $1\le i\le j\le
n$ and require computing $F(a_i,\ldots a_j)$ where $F$ is some
function of interest.

When the elements of $A$ are numbers and $F$ computes the sum of its
inputs, this problem is easily solved using linear space and constant
query time.  We create an array $B$ where $b_i$ is the sum of the
first $i$ elements of $A$.  To answer queries, we simply observe that
$a_i+\cdots+a_j=b_j-b_{i-1}$.  Indeed this approach works even if we
replace $+$ with any group operator for which each element $x$ has an
easily computable inverse $-x$.

A somewhat more difficult case is when $+$ is only a semigroup
operator, so that there is no analagous notion of $-$.  In this case,
Yao \cite{y82} shows how to preprocess a list $A$ using $O(nk)$ space
so that queries can be answered in $O(\alpha_k(n))$ time, for any
integer $k\ge 1$.  Here $\alpha_k$ is a slow growing function at the
$k$th level of the primitive recursive hierarchy.  To achieve this
result the authors show how to construct a graph $G$ with vertex set
$V=\{1,\ldots,n\}$ such that, for any pair of indices $1\le i\le j\le
n$, $G$ contains a path from $i$ to $j$ of length at most
$\alpha_k(n)$ that visits nodes in increasing order.  By labelling
each edge $(u,v)$ of $G$ with the sum of the elements
$a_u,\ldots,a_v$, queries are answered by simply summing the edge
labels along a path.  This result is optimal when $F$ is defined by a
general semigroup operator \cite{y85}.

A special case of a semigroup operator is the $\min$ (or $\max)$
operator.  In this case, the function $F$ is the function that takes
the minimum (respectively maximum) of its inputs. By making use of the
special properties of the min and max functions several researchers
\cite{bf00,bbgs89} have given data structures of size $O(n)$ that can
answer range minimum queries in $O(1)$ time.  The most recent, and
simplest, of these is due to Bender and Farach-Colton \cite{bf00}.

Range queries also have a natural generalization to trees, where they
are sometimes call \emph{path queries}.  In this setting, the input is
a tree $T$ with labels on its nodes and a query consists of two nodes
$u$ and $v$. To answer a query, a data structure must compute
$F(l_1,\ldots,l_k)$, where $l_1,\ldots,l_k$ is the set of labels
encountered on the path from $u$ to $v$ in $T$.  For group operators,
these queries are easily answered by an $O(n)$ space data structure in
$O(1)$ time using data structures for lowest-common-ancestor queries.
For semi-group operators, these queries can be answered using the same
resource bounds as for lists \cite{y82,y85}.

In this paper we consider two new types of range queries that, to the
best of our knowledge, have never been studied before.  In particular,
we consider range queries where $F$ is the function that computes a
mode or median of its input.  A mode of a multiset $S$ is an element
of $S$ that occurs at least as often as any other element of $S$.  A
median of $S$ is the element that is greater than or equal to exactly
$\floor{|S|/2}$ elements of $S$.  Our results for range mode and
range median queries are summarized in \tabref{results}.  Note that
neither of these queries is easily expressed as a group, semi-group,
or min/max query so they require completely new data structures.

\begin{table}
\begin{center}\begin{tabular}{|c|l|l|l|l|}\hline
\multicolumn{5}{|c|}{Range Mode Queries on Lists} \\ \hline
\S   & Space & Query Time & Space $\times$ Time & Restrictions \\ \hline\hline
\ref{sec:mode-tradeoff} & $O(n^{2-2\epsilon})$ & $O(n^\epsilon\log n)$ & $O(n^{2-\epsilon}\log n)$ & $0<\epsilon\le 1/2$ \\
\ref{sec:mode-fastquery} & $O(n^2\log\log n/\log n)$ & $O(1)$ & $O(n^2\log\log n/\log n)$ & -- \\ \hline
\multicolumn{5}{c}{} \\ \hline
\multicolumn{5}{|c|}{Range Mode Queries on Trees} \\ \hline
\S   & Space & Query Time & Space $\times$ Time & Restrictions \\ \hline\hline
\ref{sec:mode-tradeoff} & $O(n^{2-2\epsilon})$ & $O(n^\epsilon\log n)$ & $O(n^{2-\epsilon}\log n)$ & $0<\epsilon\le 1/2$ \\ \hline
\multicolumn{5}{c}{} \\ \hline
\multicolumn{5}{|c|}{Range Median Queries on Lists} \\ \hline
\S   & Space & Query Time & Space $\times$ Time & Restrictions \\ \hline\hline
\ref{sec:median-tradeoff-one} & $O(n\log^2 n/\log\log n)$ & $O(\log n)$ & $O(n\log^3 n/\log\log n)$ & -- \\
\ref{sec:median-fastquery} & $O(n^2\log\log n/\log^2 n)$ & $O(1)$ & $O(n^2\log\log n/\log^2 n)$ & -- \\
\ref{sec:median-tradeoff-two} & $O(n\log_b n)$ & $O(b\log^2 n/\log b)$ & $O(nb\log^3 n/\log^2 b)$ & $1\le b\le n$ \\
\ref{sec:median-tradeoff-two} & $O(n)$ & $O(n^\epsilon)$ & $O(n^{1+\epsilon})$ & $\epsilon > 0$ \\ \hline

\multicolumn{5}{c}{} \\ \hline
\multicolumn{5}{|c|}{Range Median Queries on Trees} \\ \hline
\S   & Space & Query Time & Space $\times$ Time & Restrictions \\ \hline\hline
\ref{sec:median-tree} & $O(n\log^2 n)$ & $O(\log n)$ & $O(n\log^3 n)$ & -- \\ \hline
\end{tabular}\end{center}
\caption{Summary of results in this paper.}\tablabel{results}
\end{table}

The remainder of this paper is organized as follows:  In \secref{mode-list} we
consider range mode queries on lists.  In \secref{mode-tree} we discuss 
range mode queries on trees.  In \secref{median-list} we study range median
queries on lists.  In \secref{median-tree} we present data structures for
range median queries on trees.  Finally, in \secref{conclusions} we summarize
and conclude with open problems.  

None of the four lemmas used in this paper are surprising, but some of their
proofs are rather involved.  Therefore this extended abstract omits the proofs
of all lemmas.

\section{Range Mode Queries on Lists}\seclabel{mode-list}

In this section, we consider range mode queries on an list
$A=a_1,\ldots,a_n$.  More precisely, our task is to preprocess $A$ so
that, for any indices $i$ and $j$, $1\le i\le j\le n$, we can return
an element of $a_i,\ldots,a_j$ that occurs at least as frequently as
any other element.  Our approach is to first preprocess $A$ for
\emph{range counting queries} so that, for any $i$, $j$ and $x$ we can
compute the number of occurences of $x$ in $a_i,\ldots,a_j$.  Once we
have done this, we will show how a range mode query can be answered
using a relatively small number of these range counting queries.

To answer range counting queries on $A$ we use a collection of sorted
arrays, one for each unique element of $A$.  The array for element
$x$, denoted $A_x$ contains all the indices $1\le i\le n$ such that
$a_i=x$, in sorted order.  Now, simply observe that if we search for
$i$ and $j$ in the array $A_x$, we find two indices $k$ and $l$,
respectively, such that, the number of occurences of $x$ in
$a_i,\ldots,a_j$ is $l-k$.  Thus, we can answer range counting queries
for $x$ in $O(\log n)$ time.  Furthermore, since each position in $A$
contributes exactly one element to one of these arrays, the total size
of these arrays is $O(n)$, and they can all be computed easily in
$O(n\log n)$ time.

The remainder of our solution is based on the following simple lemma about modes
in the union of three sets.

\begin{lem}\lemlabel{mode-of-three}
Let $A$, $B$ and $C$ be any multisets.  Then, if a mode of $A\cup B\cup C$ is not in
$A$ or $C$ then it is a mode of $B$.
\end{lem}

In the next two subsections we show how to use this observation to obtain
efficient data structures for range mode queries.  In the first section we show
how it can be used to obtain an efficient time-space tradeoff.  In the
subsequent section we show how to it can be used to obtain an data structure
with $O(1)$ query time that uses subquadratic space.

\subsection{A Time-Space Tradeoff}\seclabel{mode-tradeoff}

To obtain a time-space tradeoff, we partition the list $A$ into $b$
blocks, each of size $n/b$.  We denote the $i$th block by $B_i$.  For
each pair of blocks $B_i$ and $B_j$, we compute the mode $m_{i,j}$ of
$B_{i+1}\cup\cdots\cup B_{j-1}$ and store this value in a lookup table
of size $O(b^2)$.  At the same time, we convert $A$ into an array so
that we can access any element in constant time given its index.  This
gives us a data structure of size $O(n+b^2)$. 

To answer a range mode query $(i,j)$ there are two cases to consider.
In the first case, $j-i\le n/b$, in which case we can easily compute
the mode of $a_i,\ldots,a_j$ in $O((n/b)\log n)$ time by, for example,
sorting $a_i,\ldots,a_j$ and looking for the longest run of
consecutive equal elements.

The second case occurs when $j-i > n/b$, in which case $a_i$ and $a_j$
are in two different blocks (see \figref{mode-blocks}).  Let $B_{i'}$
be the block containing $i$ and let $B_{j'}$ be the block containing
$j$.  \lemref{mode-of-three} tells us that the answer to this query is
either an element of $B_{i'}$, an element of $B_{j'}$, or is the mode
$m_{i',j'}$ of $B_{i'+1}\cup\cdots\cup B_{j'+1}$.  Thus, we have a set
of at most $2n/b+1$ candidates for the mode. Using the range counting
arrays we can determine which of these candidates is a mode by
performing $2n/b+1$ queries each taking $O(\log n)$ time, for a query
time $O((n/b)\log n)$.  By setting $b=n^{1-\epsilon}$, we obtain the
following theorem:

\begin{figure}
\begin{center}\includegraphics{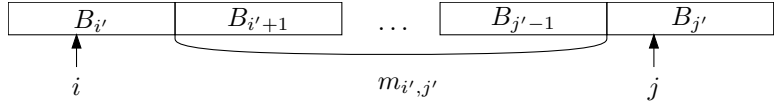}\end{center}
\caption{The mode of $a_i,\ldots,a_j$ is either an element of $B_{i'}$,
an element of $B_{j'}$ or is the mode $m_{i',j'}$ of $B_{i'+1},\ldots,B_{j'+1}$.}
\figlabel{mode-blocks}
\end{figure}

\begin{thm}\thmlabel{mode-tradeoff} For any $0<\epsilon\le 1/2$, there exists a
data structure of size $O(n^{2-2\epsilon})$ that answers range mode
queries on lists in time $O(n^\epsilon\log n)$.\footnote{The query
time of \thmref{mode-tradeoff} can be improved by observing that our
range counting data structure operates on the universe $1,\ldots,n$ so
that using complicated integer searching data structures
\cite{e77,t96,w83}, the logarithmic term in the query time can be
reduced to a doubly-logarithmic term.  We observed this but chose not
to pursue it because the theoretical improvement is negligible
compared to the polynomial factor already in the query time. The same
remarks apply to the data structure of \secref{mode-tree}}
\end{thm}

\subsection{A Constant Query-Time Subquadratic Space Solution}\seclabel{mode-fastquery}

At one extreme, \thmref{mode-tradeoff} gives an $O(n)$ space, $O(\sqrt{n}\log
n)$ query time data structure for range mode queries.  Unfortunately, at the
other extreme it gives an $O(n^2)$ space, $O(\log n)$ query time data
structure.  This is clearly non-optimal since with $O(n^2)$ space we could
simply precompute the answer to each of the $n\choose 2$ possible queries and
then answer queries in constant time.  In this section we show that it is
possible to do even better than this by giving a data structure of subquadratic
size that answers queries in constant time.

Let $k=n/b$ and consider any pair of blocks $B_{i'}$ and $B_{j'}$.
There are $k^2$ possible range mode queries $(i,j)$ such that $i$ is
in $B_{i'}$ and $j$ is in $B_{j'}$. Each such query returns a result
which is either an element of $B_{i'}$, an element of $B_{j'}$ or the
mode of $B_{i'+1}\cup\cdots\cup B_{j'+1}$. Therefore, we could store
the answers to all such queries in a table of size $k^2$, where each
table entry is an integer in the range $0,\ldots,2k$ that represents
one of these $2k+1$ possible outcomes.  The total number of such
tables is $(2k+1)^{k^2}$ and each table has size $O(k^2)$, so the
total cost to store all such tables is only $O(k^2(2k+1)^{k^2})$.
Therefore, if we choose $k=\sqrt{\log n/\log\log n}$, the total cost
to store all these tables is only $O(n^2\log\log n/\log n)$.

After computing all these tables, for each pair of blocks $B_{i'}$ and
$B_{j'}$ we need only store a pointer to the correct table and the
value of the mode $m_{i',j'}$ of $B_{i'+1}\cup\cdots\cup B_{j'-1}$.
Then, for any range mode query with endpoints in $B_{i'}$ and $B_{j'}$
we need only perform a table lookup and use the integer result to
report the mode either as an element of $B_{i'}$ an element of
$B_{j'}$ or $m_{i',j'}$.  The total size of this data structure is
$O(b^2+n)=O((n/k)^2+n)=O(n^2\log\log n/\log n)$.

To handle range mode queries $(i,j)$ where $i$ and $j$ belong to the
same block, we simply precompute all solutions to all possible queries
where $i$ and $j$ are in the same block.  The total space required for
this is $O(bk^2)=O(n\log^c n)$ which is much smaller than the space
already used.

\begin{thm}
There exists a data structure of size $O(n^2\log\log n/\log n)$ that can answer
range mode queries on lists in $O(1)$ time.
\end{thm}

\section{Range Mode Queries on Trees}\seclabel{mode-tree}

In this section we consider the problem of range mode queries on
trees.  The outline of the data structure is essentially the same as
our data structure for lists, but there are some technical
difficulties which come from the fact that the underlying graph is a
tree.  

We begin by observing that we may assume the underlying tree $T$ is a
rooted binary tree.  To see this, first observe that we can make $T$
rooted by choosing any root.  We make $T$ binary by expanding any node
with $d>2$ children into a complete binary tree with $d$ leaves.  The
root of this little tree will have the original label of the node we
expanded and all other nodes that we create are assigned unique labels
so that they are never the answer to a range mode query (unless no
element in the range occurs more than once, in which case we can
correctly return the first element of the range).  This transformation
does not increase the size of $T$ by more than a small constant
factor.

To mimic our data structure for lists we require two ingredients:
(1)~we should be able to answer range counting queries of the form:
Given a label $x$ and two nodes $u$ and $v$, how many times does the
label $x$ occur on the path from $u$ to $v$? and (2)~we must be able
to partition our tree into $O(b)$ subtrees each of size approximately
$n/b$.  

We begin with the second ingredient, since it is the easier of the
two.  To partition $T$ into subtrees we make use of the well-known
fact (see, e.g., Reference~\cite{c82}) that every binary tree has an edge
whose removal partitions the tree into two subtrees neither of which
is more than $2/3$ the size of the original tree.  By repeatedly
applying is fact, we obtain a set of edges whose removal partitions
our tree into $O(b)$ subtrees none of which has size more than $n/b$.
For each pair of these subtrees, we compute the mode of the labels on
the path from one subtree to the other and store all these modes in a
table of size $O(b^2)$.  Also, we give a new data field to each node
$v$ of $T$ so that in constant time we can determine the index of the
subtree to which $v$ belongs.

Next we need a concise data structure for answering range counting
queries.  Define the lowest-common-ancestor (LCA) of two nodes $u$ and
$v$ in $T$ to be the node on the path from $u$ to $v$ that is closest
to the root of $T$.   Let $x(v)$ denote the number of nodes labelled
$x$ on the path from the root of $T_x$ to $v$, or 0 if $v$ is nil.
Suppose $w$ is the LCA of $u$ and $v$.  Then it is easy to verify that
the number of nodes labelled $x$ on the path from $u$ to $v$ in $T$ is
exactly $x(u)+x(v)-2x(\mathrm{parent}(w))$, where $\mathrm{parent}(w)$
denotes the parent of $w$ in $T$ or nil if $w$ is the root of $T$ (see
\figref{mode-lca}).  

\begin{figure}
\begin{center}\includegraphics{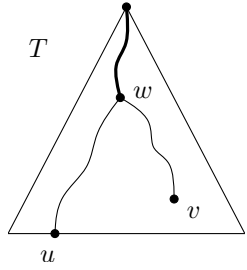}\end{center}
\caption{The number of nodes labelled $x$ on the path from $u$ to $v$
is $x(u)+x(v)-2x(\mathrm{parent}(w))$.}\figlabel{mode-lca}
\end{figure}

There are several data structures for preprocessing $T$ for LCA
queries that use linear space and answer queries in $O(1)$ time. Thus
all that remains is to give a data structure for computing $x(u)$ for
any value $x$ and any node $u$ of $T$.  Consider the minimal subtree
of $T$ that is connected and contains the root of $T$ as well as all
nodes whose label is $x$.  Furthermore, contract all degree 2 vertices
in this subtree with the exception of the root and call the resulting
tree $T_x$ (see \figref{mode-intervals}). It is clear that the tree
$T_x$ has size proportional to the number of nodes labelled $x$ in the
original tree.  Furthermore, by preprocessing $T_x$ with an LCA data
structure and labelling the nodes of $T_x$ with their distance to
root, we can compute, for any nodes $u$ and $v$ in $T_x$, the number
of nodes labelled $x$ on the path from $u$ to $v$ in $T$.

The difficulty now is that we can only do range counting queries
between nodes $u$ and $v$ that occur in $T_x$ and we need to answer
these queries for any $u$ and $v$ in $T$.  What we require is a
mapping of the nodes of $T$ onto corresponding nodes in $T_x$. More
precisely, for each node $v$ in $T$ we need to be able to identify the
first node labelled $T_x$ encountered on the path from $v$ to the root
of $T$.  Furthermore, we must be able to do this with a data structure
whose size is related to the size of $T_x$, not $T$.

To achieve this mapping, we perform an \emph{interval labelling} of
the nodes in $T$ (see \figref{mode-intervals}): We label the nodes of
$T$ with consecutive integers by an in-order traversal of $T$.  With
each internal node $v$ of $T$, we assign the minimum interval that
contains all of the integer labels in the subtree rooted at $v$.  Note
that every node in $T_x$ is also a node in $T$, so this also gives an
interval labelling of the corresponding nodes in $T_x$ (although the
intervals are not minimal).  Consider a node $v$ of $T$ whose integer
label is $g$.  Then it is easy to verify that the first node labelled
$x$ on the path from $v$ to the root of $T$ is the node of $T_x$ with
the smallest interval label that contains $g$.  Next, observe that if
we sort the endpoints of these intervals then in any subinterval
defined by two consecutive endpoints the answer to a query is the
same.  Therefore, by sorting the endpoints of the intervals of nodes
in $T_x$ and storing these in a sorted array we can answer these
queries in $O(\log n)$ time using a data structure of size $O(|T_x|)$.

\begin{figure}
\begin{center}
\begin{tabular}{c@{\hspace{1.5cm}}c}
\includegraphics{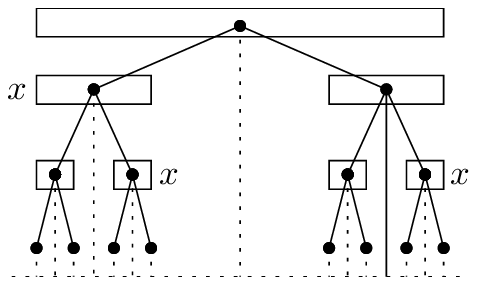} & \includegraphics{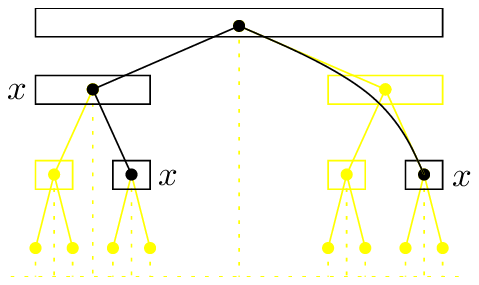} \\
$T$ & $T_x$
\end{tabular}
\end{center}
\caption{The trees $T$ and $T_x$ and their interval labelling.}
\figlabel{mode-intervals}
\end{figure}

To summarize, we have described all the data structures needed to
answer range counting queries in $O(\log n)$ time using a data
structure of size $O(n)$.  To answer a range mode query $(u,v)$ we
first lookup the two subtrees $T_u$ and $T_v$ of $T$ that contain $u$
and $v$ as well as a mode $m_{u,v}$ of all the labels encountered on
the path from $T_u$ to $T_v$.  We then perform range counting queries
for each of the distinct labels in $T_u$ and $T_v$ as well as
$m_{u,v}$ to determine an overall mode.  The running time and storage
requirements are identical to the data structure for lists.

\begin{thm}
For any $0<\epsilon\le 1/2$, there exists a data structure of size
$O(n^{2-2\epsilon})$ that answers range mode queries on trees in
$O(n^\epsilon\log n)$ time.
\end{thm}

\section{Range Median Queries on Lists}\seclabel{median-list}

In this section we consider the problem of answering range median
queries on lists.  To do this, we take the same general approach used
to answer range mode queries.  We perform a preprocessing of $A$ so
that our range median query reduces to the problem of computing the
median of the union of several sets.

\subsection{The Median of Several Sorted Sets}

In this section we present three basic results that will be used in
our range median data structures.  

An \emph{augmented} binary search tree is a binary search tree in
which each node contains a \emph{size} field that indicates the number
of nodes in the subtree rooted at that node. This allows, for example,
determining the rank of the root in constant time (it is the size of
the left subtree plus 1) and indexing an element by rank in $O(\log
n)$ time.  Suppose we have three sets $A$, $B$, and $C$, stored in
three augmented binary search trees $T_A$, $T_B$ and $T_C$,
respectively, and we wish find the element of rank $i$ in $A\cup B\cup
C$.  The following lemma says that we can do this very quickly.

\begin{lem}\lemlabel{find-median-of-three}
Let $T_A$, $T_B$, and $T_C$ be three augmented binary search trees on the sets
$A$, $B$, and $C$, respectively.  There exists an $O(h_A+h_B+h_C)$ time
algorithm to find the element with rank $i$ in $A\cup B\cup C$, where $h_A$,
$h_B$ and $h_C$ are the heights of $T_A$, $T_B$ and $T_C$, respectively.
\end{lem}

Another tool we will make use of is a method of finding the median in
the union of many sorted arrays. 

\begin{lem}\lemlabel{find-median-of-many}
Let $A_1,\ldots,A_k$ be sorted arrays whose total size is $O(n)$.  There exists
an $O(k\log n)$ time algorithm to find the element with rank $i$ in 
$A_1\cup\cdots\cup A_k$.
\end{lem}

Finally, we also make use of the following fact which plays a role
analagous to that of \lemref{mode-of-three}.

\begin{lem}\lemlabel{median-of-three}
Let $A$, $B$, and $C$ be three sets such that $|A|=|C|=k$ and
$|B|>4k$.  Then the median of $A\cup B\cup C$ is either in $A$, in $C$
or is an element of $B$ whose rank in $B$ is in the range
$[|B|/2-2k,|B|/2+2k]$.
\end{lem}

\subsection{A First Time-Space Tradeoff}\seclabel{median-tradeoff-one}

To obtain our first data structure for range median queries we proceed
in a manner similar to that used for range mode queries.  We partition
our list $A$ into $b$ blocks $B_1,\ldots,B_{n/b}$ each of size $n/b$.
We will create two types of data structures.  For each block we will
create a data structure that summarizes that block.  For each pair of
blocks we will create a data structure that summarizes all the
elements between that pair of blocks.

To process each block we make use of \emph{persistent augmented binary
search trees}.  These are search trees in which, every time an item is
inserted or deleted, a new \emph{version} of the tree is created.
These trees are called persistent because they allow accesses to all
previous versions of the tree.  The simplest method of implementing
persistent augmented binary search trees is by \emph{path-copying}
\cite{km83,m82,m84,rtd83,s85}.  This results in $O(\log n)$ new nodes
being created each time an element is inserted or deleted, so a
sequence of $n$ update operations creates a set of $n$ trees that are
represented by a data structure of size $O(n\log
n)$.\footnote{Although there are persistent binary search trees that
require only $O(n)$ space for $n$ operations \cite{dsst89,st86}, these
trees are not \emph{augmented} and thus do not work for our
application.  In particular, they do not allow us to make use of
\lemref{find-median-of-three}.}

For each block $B_{i'}=b_{i',1},\ldots,b_{i',n/b}$, we create two
persistent augmented search trees $\overrightarrow{T}_{i'}$ and
$\overleftarrow{T}_{i'}$.  To create $\overrightarrow{T}_{i'}$ we
insert the elements $b_{i',1},b_{i',2},\ldots,b_{i',n/b}$ in that
order.  To create $\overleftarrow{T}_{i'}$ we insert the same elements
in reverse order, i.e., we insert
$b_{i',n/b},b_{i',n/b-1},\ldots,b_{i',1}$.   Since these trees are
persistent, this means that, for any $j$, $1\le j\le n/b$, we have
access to a search tree $\overrightarrow{T}_{i',j}$ that contains
exactly the elements $b_{i',1},\ldots,b_{i',j}$ and a search tree
$\overleftarrow{T}_{i',j}$ that contains exactly the elements
$b_{i',j},\ldots,b_{i',n/b}$.

For each pair of blocks $B_{i'}$ and $B_{j'}$, $1\le i'<j'\le n$, we
sort the elements of $B_{i'+1}\cup\cdots\cup B_{j'-1}$ and store the
elements whose ranks are within $2n/b$ of the median in a sorted array
$A_{i',j'}$.  Observe that, by \lemref{median-of-three}, the answer to
a range median query $(i,j)$ where $i=i'n/b+x$ is in block $i'$ and
$j=j'n/b+y$ is in block $j'$, is in one of $\overleftarrow{T}_{i',x}$,
$A_{i',j'}$ or $\overrightarrow{T}_{j',y}$ (see
\figref{median-blocks}).  Furthermore, given these two trees and one
array, \lemref{find-median-of-three} allows us to find the median in
$O(\log n)$ time. 

\begin{figure}
\begin{center}\includegraphics{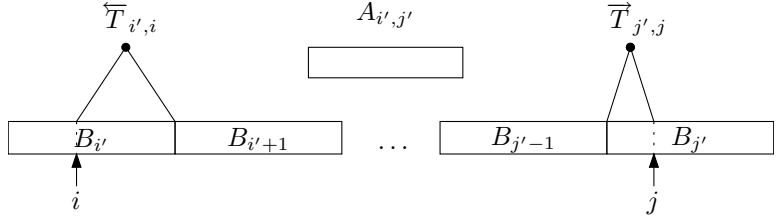}\end{center}
\caption{The median of $a_i,\ldots,a_j$ can be computed 
from two persistent search trees.} 
\comment{$\overleftarrow{T}_{i',x}$, 
$A_{i',j'}$ and 
$\overrightarrow{T}_{j',y}$ in $O(\log n)$ time.
}
\figlabel{median-blocks}
\end{figure}

Thus far, we have a data structure that allows us to answer any range
median query $(i,j)$ where $i$ and $j$ are in different blocks $i'$
and $j'$.  The size of the data structure for each block is
$O((n/b)\log n)$ and the size of the data structure for each pair of
blocks is $O(n/b)$.  Therefore, the overall size of this data
structure is $O(n(b+\log n))$.  To obtain a data structure that
answers queries for \emph{any} range median query $(i,j)$ including
$i$ and $j$ in the same block, we build data structures recursively
for each block.  The size of all these data structures is given by the
recurrence \[ T_n = bT_{n/b} + O(n(b+\log n)) = O((n(b+\log n))\log_b
n \enspace .\]

\begin{thm}
For any $1\le b\le n$, there exists a data structure of size $O(n(b+\log n)\log_b n$
that answers range median queries on lists in time $O(\log (n/b))$.
\end{thm}

At least asymptotically, the optimal choice of $b$ is $b=\log n$. In
this case, we obtain an $O(n\log^2 n/\log\log n)$ space data structure
that answers queries in $O(\log n)$ time.  In practice, the choice
$b=2$ is probably preferable since it avoids having to compute the
$A_{i',j'}$ arrays altogether and only ever requires finding the
median in two augmented binary search trees.   The cost of this
simplification is only an $O(\log\log n)$ factor in the space
requirement.

\subsection{A Constant Query Time Subquadratic Space Data Structure}\seclabel{median-fastquery}

Next we sketch a range median query data structure with constant query
time and subquadratic space.  The data structure is essentially the
same as the range mode query data structure described in
\secref{mode-fastquery} modified to perform median queries.  The
modifications are as follows:  For each pair of blocks $B_{i'}$ and
$B_{j'}$ we need only consider the set of $6k$ elements that are
potential medians of queries with endpoints $i$ and $j$ in $B_{i'}$
and $B_{j'}$.  We can also create a normalized version of these
elements, so that each element is a unique integer in the range
$1,\ldots,6k$.  In this way, we only need to create $(6k)!$ different
lookup tables, each of size $O(k^2)$.

To summarize, storing all the lookup tables takes $O(k^2(6k)!)$ space.
For each pair of blocks we must store a pointer to a lookup table as
well as an array of size $6k$ that translates ranks in the lookup
table to elements of $A$, for a total space of $O(b^2k)$.  For each
block we precompute and store all the solutions to queries with both
endpoints in that block.  Setting $k=c\log n/\log\log n$ for
sufficiently small $c$, we obtain an overall space bound of
$O(n^2\log\log n/\log^2 n)$.

\begin{thm}
There exists a data structure of size $O(n^2\log\log n/\log^2 n)$ that can
answer range median queries on lists in $O(1)$ time.
\end{thm}

\subsection{A Data Structure Based on Range Trees}\seclabel{median-tradeoff-two}

Next we describe a range median data structure based on the same
principle as Lueker and Willard's range trees \cite{l78,w85}.  This
data structure stores $a_1,\ldots,a_n$ at the leaves of a complete
$b$-ary tree $T$ in the order in which they appear in $A$.  At each
internal node $v$ of this tree we keep a sorted array containing all
the elements of $A$ that appear at leaves in the subtree rooted at
$v$.  It is clear that this tree, including the arrays stored at all
the nodes, has size $O(n\log_b n)$.

To use this tree to answer a range query $(i,j)$, consider the two
paths $P_i$ and $P_j$ from the root of $T$ to the leaf containing
$a_i$ and the leaf containing $a_j$, respectively (see
\figref{median-range}).  These two paths share some nodes for a period
of time and then diverge.  Observe that, after this point, by looking
at the sorted arrays at nodes to the right of $P_i$ and to the left of
$P_j$ we obtain a partition of $a_i,\ldots,a_j$ into a set of sorted
arrays.  The number of these arrays is at most $b\log_b n$ and their
total size is at most $n$.  Therefore, by \lemref{find-median-of-many}
we can answer the range median query $(i,j)$ in $O(b\log^2 n/\log b)$
time. 

\begin{figure}
\begin{center}\includegraphics{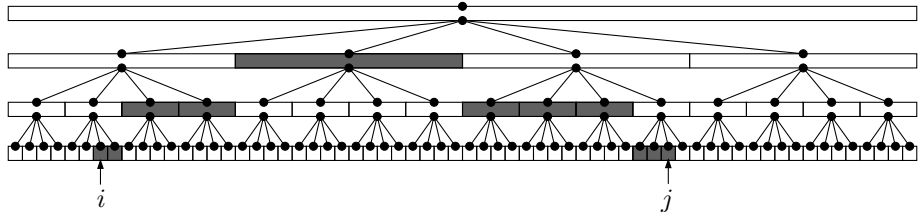}\end{center}
\caption{Using range trees to perform range median queries. The median of $a_i,\ldots,a_j$ is the median of the elements in the $O(b\log_b n)$ shaded arrays.}
\figlabel{median-range}
\end{figure}

\begin{thm} 
For any integer $1\le b\le n$, there exists a data structure of size
$O(n\log_b n)$ size that answers range median queries on lists in
$O(b\log^2 n/\log b)$ time.  In particular, for any constant
$\epsilon>0$ there exists a data structure of size $O(n)$ that answers
range median queries in $O(n^\epsilon)$ time.  
\end{thm}

\section{Range Median Queries on Trees}\seclabel{median-tree}

Next we consider how to answer range median queries on trees.  As
before, we may assume that $T$ is a binary tree by converting nodes
node with $d>2$ children into complete binary trees.  In these little
trees we subdivide edges to ensure that the number of internal nodes
in any root to leaf path is even and label these nodes alternately
with $-\infty$, $+\infty$ so as not affect the median on any path
between two of the original nodes of $T$.

Our method is simply the binary version of the basic method in
\secref{median-tradeoff-one} for lists.  We first find a centroid edge
$(a,b)$ of $T$ whose removal partitions $T$ into two subtrees $T_a$
and $T_b$ each of size at most $2/3$ the original size of $T$.  For
each node $u$ in $T_a$, we would like to have access to an augmented
search tree that contains exactly the labels on the path from $u$ to
$a$.  To achieve this, we proceed as follows: To initialize the
algorithm we insert the label of $a$ into a persistent augmented
binary search tree, mark $a$ and define this new tree to be the
\emph{tree of $a$}.  While some marked node $u$ of $T_a$ has an
unmarked child $v$, we insert the label of $v$ into the tree of $u$,
mark $v$, and define this new tree to be the tree of $v$.  Note that
because we are using persistent search trees, this leaves the tree of
$u$ unchanged.  In this way, for any node $u$ in $T_a$, the tree of
$u$ contains exactly the labels of nodes on the path from $u$ to $a$.
We repeat the same procedure for $T_b$, and this creates a data
structure of size $O(n\log n)$.

To answer a range median query $(u,v)$ where $u$ is in $T_a$ and $v$
is in $T_b$, we only need to find the median of all labels stored in
the tree of $a$ and the tree of $b$.  By \lemref{find-median-of-three}
this can be done in $O(\log n)$ time.  To answer range median queries
$(u,v)$ where both $u$ and $v$ are in $T_a$ (or $T_b$) we recursively
build data structures for range median queries in $T_a$ and $T_b$.
The total size of all these data structures is \[  T_n = T_{\alpha n}
+ T_{(1-\alpha)n} + O(n\log n) = O(n\log^2 n) \enspace ,\] where
$1/3\le\alpha\le 2/3$ and they can answer range median queries in
$O(\log n)$ time.

\begin{thm}\thmlabel{median-tree}
There exists a data structure of size $O(n\log^2 n)$ that can answer
range median queries in trees in $O(\log n)$ time. 
\end{thm}

It is tempting to try and shave a $\log\log n$ factor off the storage
requirement of \thmref{median-tree} by using a $\log n$-ary version of
the above scheme as we did in \secref{median-tradeoff-one}.  However,
the reason this worked for lists is that, for any block, a query
either extends to the left or right boundary of that block, so only
two persistent search trees are needed.  However, if we try to make a
$\log n$-ary partition of a tree we find that each subtree (block) can
have $\Omega(\log n)$ vertices that share an edge with another
subtree, which would require $\Omega(\log n)$ persistent search trees
per subtree.

\section{Summary and Conclusions}\seclabel{conclusions}

We have given data structures for answering range mode and
range median queries on lists and trees.  To the best of our
knowledge, we are the first to study these problems.  These problems
do not seem to admit the same techniques used to develop optimal data
structures for range queries involving group or semigroup operators. 

Essentially every result in this paper is an open problem.  There are
no lower bounds for these problems and it seems unlikely that any of
our data structures are optimal.  Thus, there is still a significant
amount of work to be done on these problems, either by improving these
results and/or showing non-trivial lower bounds for these data
structures.

\section*{Acknowledgement}

The third author would like to thank Stefan~Langerman for helpful discussions.

\bibliographystyle{plain}
\bibliography{rmq}

\comment{
\appendix
\section{Proofs of Lemmas}\lemlabel{proofs}

\begin{proof}[Proof of \lemref{mode-of-three}]
Suppose, for the sake of contradiction, that the mode $x$ of $A\cup B\cup C$
does not appear in $A$ or $C$ and is not a mode of $B$.  But then there is
a mode $y\in B$ of $B$ that occurs more frequently in $B$ than $x$.  Since
$x$ does not occur in $A$ or $C$ then $y$ also occurs more frequently in $A\cup
B\cup C$ than $x$, so $x$ is not a mode of $A\cup B\cup C$, a contradiction.
\end{proof}

\begin{proof}[Proof of \lemref{find-median-of-three}]
Trivial.
\end{proof}

\begin{proof}[Proof of \lemref{find-median-of-many}]
Trivial.
\end{proof}

\begin{proof}[Proof of \lemref{median-of-three}]

\end{proof}
}

\end{document}